\documentclass[twocolumn]{article}

\usepackage[english]{babel}
\usepackage[utf8x]{inputenc}
\usepackage[T1]{fontenc}
\usepackage{amsfonts}
\usepackage{mathrsfs}

\usepackage[a4paper,top=3cm,bottom=2cm,left=2cm,right=2cm]{geometry}

\usepackage{amsmath}
\usepackage{graphicx}
\usepackage[colorinlistoftodos]{todonotes}
\usepackage[colorlinks=true, allcolors=blue]{hyperref}
\usepackage{float}
\title{
	\usefont{OT1}{bch}{b}{n}
	\normalfont \normalsize \textsc{} \\ [8pt]
	\huge 
	Empirical Conditional Method: A New Approach to Predict Throughput in TCP Mobile Data Network\\
}
\usepackage{authblk}
\author[1]{Weijia Zheng}
\affil[1]{Department of Information Engineering, CUHK \authorcr
wjzheng@link.cuhk.edu.hk}

\date{}
\begin{document}

\maketitle

\begin{abstract}
Experience of live video streaming can be improved 
if future available bandwidth can be predicted more accurately 
at the video uploader side. 
Thus follows a natural question which is 
how to make predictions both easily and precisely in an ever-changing network. 
Researchers have developed many prediction algorithms in the literature, 
from where a simple algorithm, Arithmetic Mean (AM), stands out. 
Based on that, 
we are purposing a new method called Empirical Conditional Method (ECM) 
based on a Markov model,
hoping to utilize more information in the past data 
to get a more accurate prediction without loss of practicality. 

Through simulations, we found that 
our ECM algorithm performs better than the commonly used AM one, in the sense 
of reducing the loss by about $10\%$ comparing with AM. 
Besides, ECM also has a higher utilization rate of available bandwidth, 
which means ECM can send more data out 
while not having higher loss rate or delay, especially under a low FPS setting.
ECM can be more helpful for those who have relatively 
limited networks to reach a more considerable 
a balance between frame loss rate and video quality hence improve the quality 
of experience.

\end{abstract}

~\


\section{Introduction}

With the development of smartphones and high-speed mobile data networks such as 3G and 4G/LTE, 
live video streaming has long become part of our lives in the entertainment or casual fields. 
In recent years, it has played a vital role in the workplace as well. 
As we all experienced in person, 
the past year of 2020 witnessed Zoom's significant increase [1] in usage of remote work, 
distance education, and online social relations. 

Given the importance of video streaming and the high peak bandwidth nowadays in mobile data networks, 
bandwidth fluctuation remains a challenging obligation 
for its unpredictable characteristic by its wireless nature, 
which may affect the quality of clients' experience. 
Several prediction methods were established in the literature  
and Arithmetic Mean (AM) algorithm stands out [2] as one of 
the most suitable ones for its simplicity and fairly-good performance. 

Motivated by similar ideas of using conditional probability for prediction in other fields [3], 
this study proposes a new approach of throughput prediction,
called the Empirical Conditional Mean (ECM),
hoping the empirical conditional probability distribution 
from past data to contribute more informative suggestions 
when predicting future throughputs and thus returns a better prediction value. 

We also coded a simulator to simulate the uplink part of a 
video streaming process to test our hypothesis's validity 
and feasibility in comparison with the AM algorithm. 
The simulator uses trace-driven simulations with throughput 
trace data measured from real-world uplink network sources of 3HK 4G.

\section{Problem Statement}
In our study, to concentrate on the throughput prediction part, 
we did not consider the downlink part of the streaming process and 
did not distinguish frame types either. 
We considered the uplink part exclusively with some further simplifications. 
The following is the setup of our problem.

An uploader generates video frames one by one with an equal time difference (i.e., 1/FPS second) 
in between. It adopts a Stop-And-Wait scheme in sending frames.
Namely, it starts a new transmission process of a frame only when it knows the previous transmission finishes. 
Moreover, it always sends the newest-generated frame out. 
Should there be any not-ever-sent frame when the streaming process terminates, 
we count it as a loss. 
We calculate the loss rate by dividing the loss number by the total number of frames the uploader generated.

Each time when the uploader finished up-linking a frame $i$, it will calculate the (just now) average throughput using $C_i = s_i / t_i,$ where $s_i$ is frame $i$’s size which is determined by the uploader itself, and $t_i$ is how much time the uploader cost to up-link frame $i$, 
which is the uploader can always make a record on. When frame $i+1$ needs to be transmitted, the uploader will predict $C_{i+1}$ by $$\widehat{C_{i+1}}=g(\{ C_{j \leq i} \}),$$ where $g$ is some predicting algorithm. 

A study carried by Zou et al. [4] revealed that the optimal case happens when we can achieve 
$$C_{i+1} = \widehat{C_{i+1}},$$ thus we can encode $s_{i+1}$ such that the uploader will know frame $i+1$’s transmission is finished at the exact time when frame $i+2$ is just generated. By this, we are sending as large a frame size as possible while not losing any single frame. In consideration of reality, we added a constrain that each frame should have a minimal frame size because we cannot let frame size be arbitrarily small. 

Under the above settings, we can then use two simple metrics to measure how an algorithm performs during a period. These are (1) loss rate and 
(2) sum of frames’ sizes (a.k.a. $\sum_{i} s_i$). We would examine how AM and ECM perform under the same network environment.

\begin{figure}[H]
	  \includegraphics[width=0.48\textwidth]{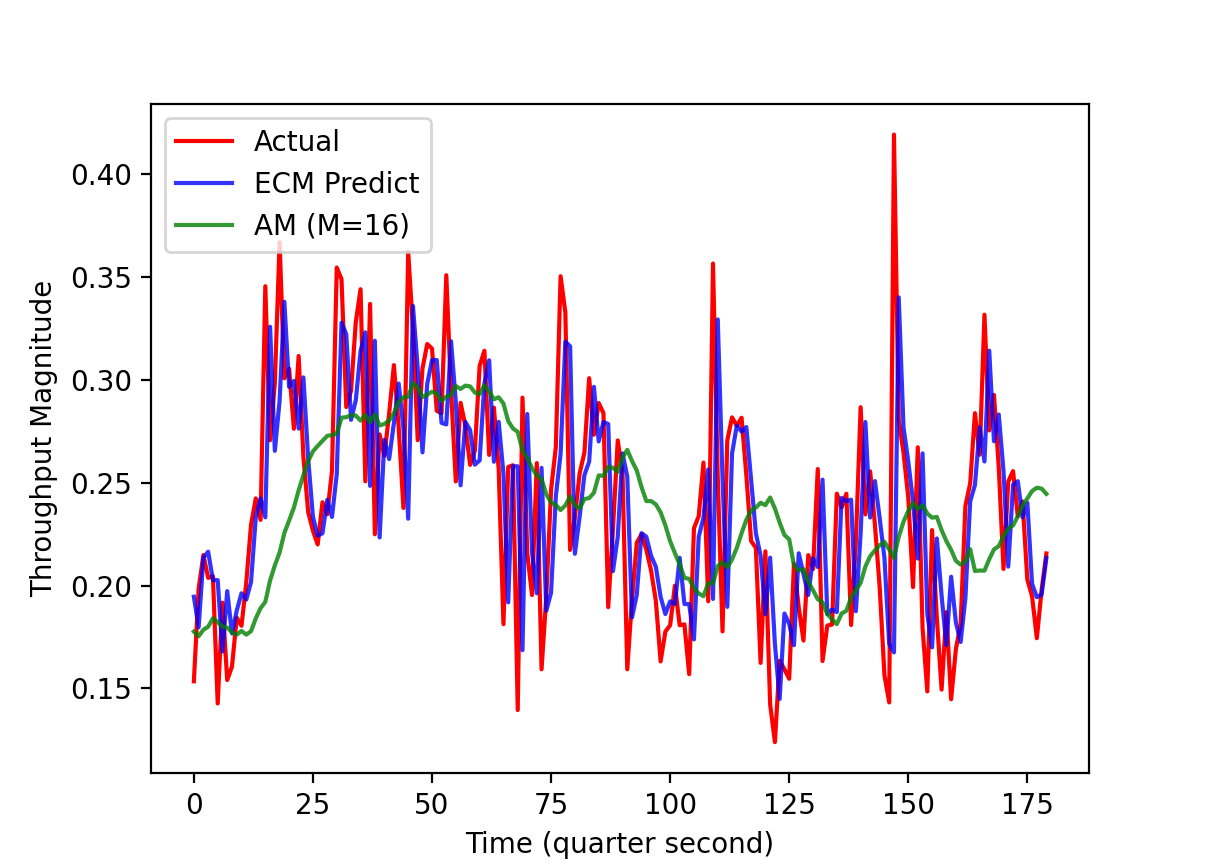}
	  \caption{Example of ECM prediction comparing with AM (M=16).}
\end{figure}

\section{AM and ECM}
\subsection{Arithmetic Mean (AM)}
The AM algorithm returns a moving average of the past $C_i$'s. 
Suppose $M$ is the number of past $C_i$'s we look back, 
then for estimating $C_{i+1},$ we calculate
$$\widehat{C_{i+1}} = \frac{1}{M}\sum_{j=1}^{M}C_{i+1-j}$$ as the point estimator.
This algorithm does not require training. Hence it is simple to apply. 
However, according to literature and simulations in our study, 
it did perform fairly well and got proposed [2][5]. 

\subsection{Empirical Conditional Mean (ECM)}
Unlike AM, Empirical Conditional Mean (ECM) does require training. 
More precisely, 
it requires some time to accumulate historical data to 
establish an empirical conditional probability space to a specific scale. 
The following shows the procedures to achieve this.

\subsubsection*{Data Binning}
Since all measured value $C_i$’s are distributed in a continuous interval on
$\mathbb{R}$, 
thus requires us to quantize/discretize the interval into several bins. 
Each bin is a state of this Markov model.
Let the minimal value possible for $C_i$’s be $L$, 
and the maximal value be $U$. 
Inside $(L,U]$, 
we discretize them into $K$ non-overlapping subintervals. Namely, 
$(L,U]=\cup_{k=1}^{K} (L_k,L_{k+1}],$ where $L_k$ denotes the minimal value in 
the $k$-th interval with $L_1=L, L_{K+1}=U$.

\subsubsection*{Generating the Probability Space}
We maintain a $K \times K$ matrix $\mathcal{S}$, 
initializie all its entries to be 0.
Every time we measured two consecutive values $C_j, C_{j+1}$ belongs to the 
$m$-th and $n$-th interval respectively, we add $\mathcal{S}_{mn}$ by 1. 

Keep this process for a period of time, 
$\mathcal{S}_{mn}$ then denotes how many times 
does a pair of two consecutive measured throughputs with 
a older value belongs to the $(L_n,L_{n+1}]$ 
followed by a newer one belongs to the $(L_m,L_{m+1}]$. 

\subsubsection*{Point Estimate}
After the training process, for estimating
$C_{i+1},$ given that $C_{i}\in (L_k, L_k+1]$ for some $k$, we calculate 

$$\widehat{C_{i+1}} = \sum_{j=1}^{K} \frac{S_{kj}}{S_{k+}}\cdot A_j,$$ 
where $S_{k+} = \sum_{i=1}^{K}S_{ki}$, and $A_j$ is an arbitary value with 
$A_j \in (L_j,L_{j+1}].$
Note that here the $S_{km}/S_{k+}$ term can be regarded as an empirical probability of 
$$Pr\{ C_{i+1}\in (L_m,L_{m+1}] ~ | ~ C_{i}\in (L_k,L_{k+1}] \}.$$ 
Instead of choosing a value with the maximal transition probability, 
we instead use an expectation value as a point estimator based on our 
simulation results.

\subsubsection*{Diminishing Mutual Information}
“Intuitively, we expect more recent past throughput data to exhibit higher correlation 
with future throughput and is thus more valuable in throughput prediction.” 
The study conducted by Liu verified (in Fig. 2 of [2]) that the mutual information between 
$C_{i}$ and $C_{i-k}, (k\in \mathbb{Z}^+,k\geq 1)$ 
tends to be non-increasing as $k$ increases. 

Note that in ECM, 
the $\mathcal{S}$ matrix can be changed to a transition matrix 
by dividing every entry by its corresponding row total. Namely, 
$[\mathcal{S_M}]_{ij}=[\mathcal{S}]_{ij}/\mathcal{S}_{i+}$ is a Markov matrix 
whose each row sums up to one. Thus regards $C_i$ as 
random variables in a Markov process [6], i.e.,
$$C_{i-2} \to C_{i-1} \to C_{i}.$$ 
According to the data processing inequality [6], 
$$I(C_{i};C_{i-1})\geq I(C_i;C_{i-2}),$$ 
which is consistent with Liu's observation result in [2]. 

\begin{figure}[H]
	  \includegraphics[width=0.48\textwidth]{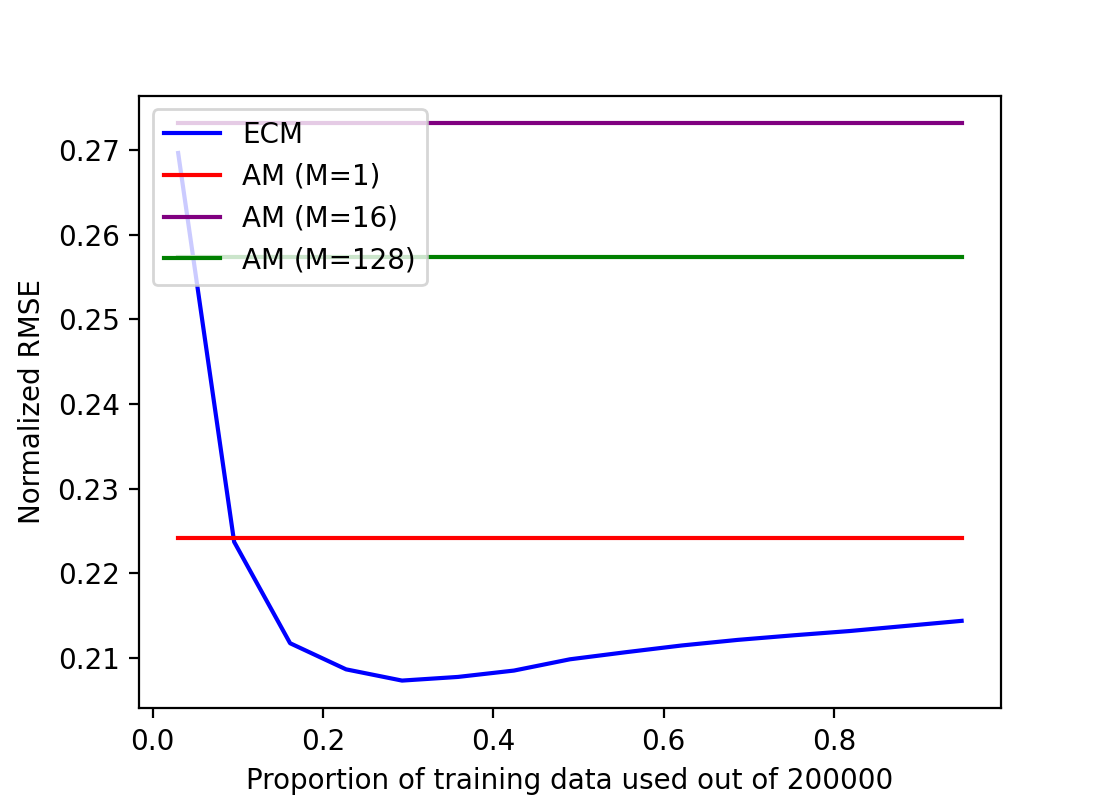}
	  \caption{Loss value of ECM turns to increase after we increasing the proportion of training data.}
\end{figure}

Nevertheless, in our experiments, this effect appears occasionally. 
As one can see in Figure 2, though overall ECM performs better than AMs, 
ECM's loss function (normalized RMSE given by [7], 
NRMSE$~=\frac{\sqrt{\frac{1}{N}\sum_{j=1}^N(C_{j}-\widehat{C_{j}})^2}}{E[C_i]}$) 
takes turn to increase as we keep feeding our matrix $\mathcal{S}$ with more data earlier and earlier back from the 
time spot we predict the future throughput. This suggests that data from a long time ago fed in may not be helpful to our predicting model. 
Therefore we set a upper bound $M_{S}$ for the grandsum of $\mathcal{S}$. Namely, $$\sum_{r=1}^{K}\sum_{c=1}^{K}\mathcal{S}_{rc} \leq M_{S}.$$ Once the grandsum exceeds $M_{S},$ we subtract 1 from the entry where an 1 is added $M_{S}$ steps ago. Then the $\mathcal{S}$ will have its grandsum being constant as $M_{S},$ in a fashion of adding 1 in some entry and subtracting 1 in (maybe) another entry in an iteration.
By doing this, we let inside our $\mathcal{S}$ there are only those most recent, thus expected-to-be the most informative and helpful data. However, do earlier data provide much less mutual information than newer ones, their correlation is not negligible even for data collected as early as 300 seconds ago [2]. 

This forgetting scheme turns out to be effective in practice, which is shown in Figures 3, 4, and 5. 
The ECMs adapted with a forgetting scheme have more total data sent, meaning utilizing the available bandwidth more than those without a forgetting scheme while having similar loss rates.

\subsubsection*{Interval Estimate}
Unlike other models that may need other statistical assumptions or 
to do regressions to achieve an confidence interval of the point estimate. 
We have already constructed a probability space by the nature of our method. 
Suppose we want to construct a $(1-\alpha)\cdot  100\%$ C.I. for $C_{i+1}$, 
given $C_{i}\in (L_m,L_{m+1}].$ Then we can take out the $m-$th row from 
$\mathcal{S_M},$ which is a discrete probability distribution having 
a similar shape with those in Figure 6 or Figure 7. Then we trim out the left and 
right tails such that the remaining probabilities sum up to be $(1-a)\cdot 100\%$.



\section{Simulation Results}
\subsection{Interval Estimate Result}
We test in two different trace data set, for both
we use 5000 data to train the model and the rest 5000 to test. 
Set the significance level to be $\alpha$ and construct a 
$(1-\alpha)\cdot 100\%$ confidence interval using the above way. 
We then would count how many observations in testing set are 
inside the confidence interval we made up with empirical conditional probability.

For data set 1, we set $\alpha=0.05$ hence suppose to hava a $95\%$ confidence 
interval, the result turns out that $95.72\%$ values in the testing set are inside 
the confidence interval. 
Similarly, for data set 2, we set $\alpha=0.2$ 
hence suppose to hava a $80\%$ confidence 
interval, the result turns out to be $80.28\%$. 
Hence experiments show that the confidence interval works
under different settings of the significance level $\alpha$.

\subsection{Point Estimate Result}
In this section, we use the point estimate as the prediction and compare with AM 
algorithm in the simulator.
As AM algorithm has a parameter $M$, the size of its sliding window, 
we let $M=1,16$ and $128$, to see AM's performance under different levels of $M$. Besides, as mentioned above, we always set a minimum frame size for each generated frame. Hence we would see how the two performance metrics, (1) loss rate and (2) sum of frames' sizes, behave against an increasing minimal frame size. Moreover, we will only inspect the cases when the loss rate is no larger than 2\% because too much loss rate at the uploader side will certainly be unacceptable.

\begin{figure}[H]
		\includegraphics[width=0.48\textwidth]{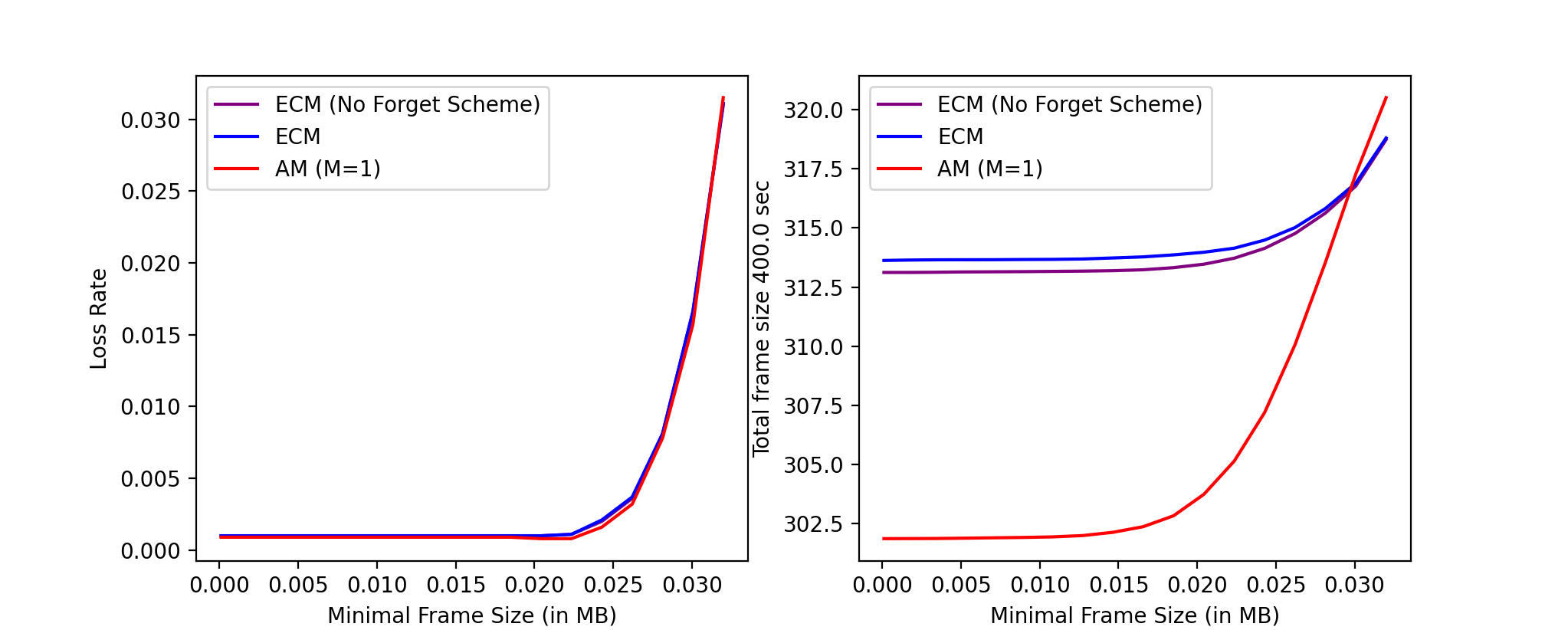}
  		\caption{Comparing with AM (M=1), ECM has slightly less loss rate and higher sent frame sizes.}
\end{figure}

\begin{figure}[H]
	\includegraphics[width=0.48\textwidth]{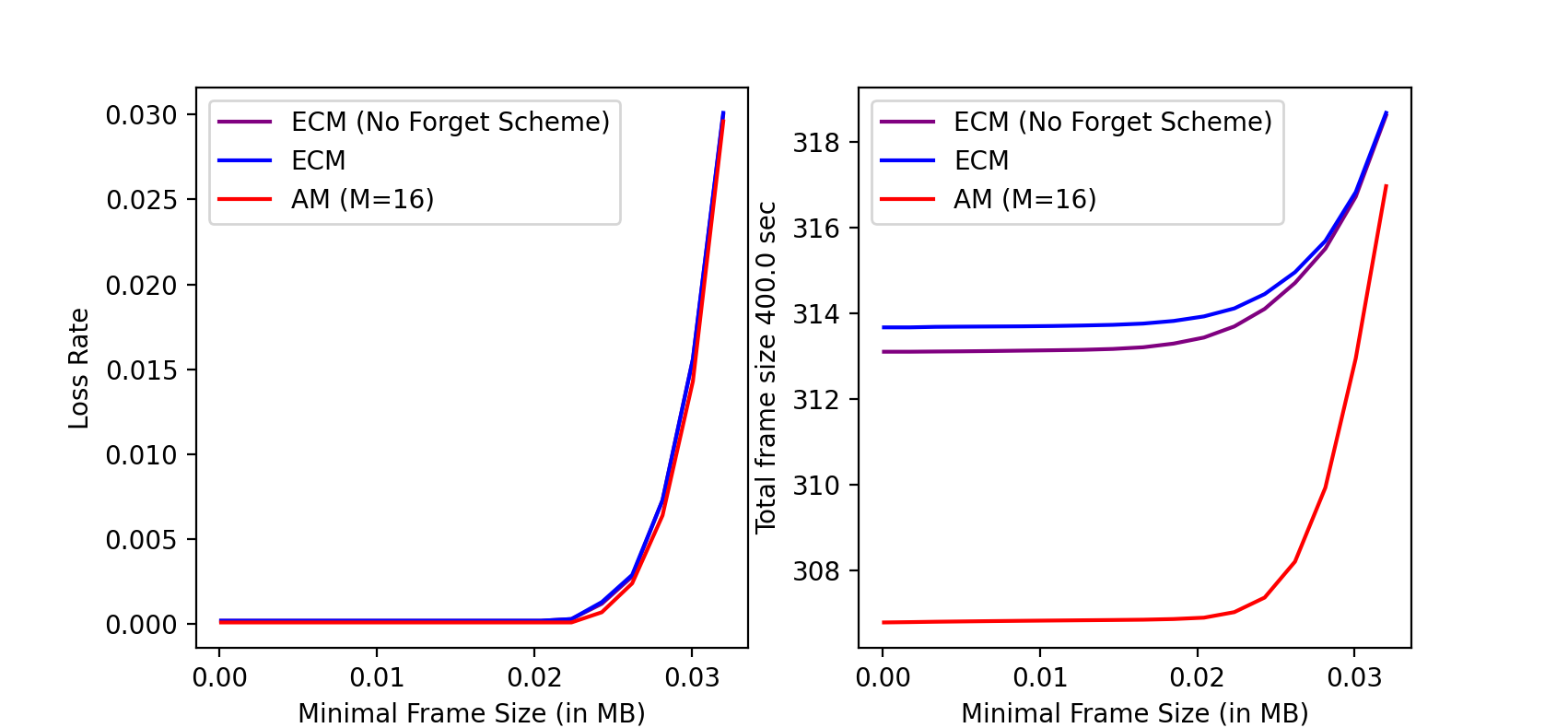}
	  \caption{Comparing with AM (M=16), ECM has similar loss rate and higher sent frame sizes.}
\end{figure}

\begin{figure}[H]
	\includegraphics[width=0.48\textwidth]{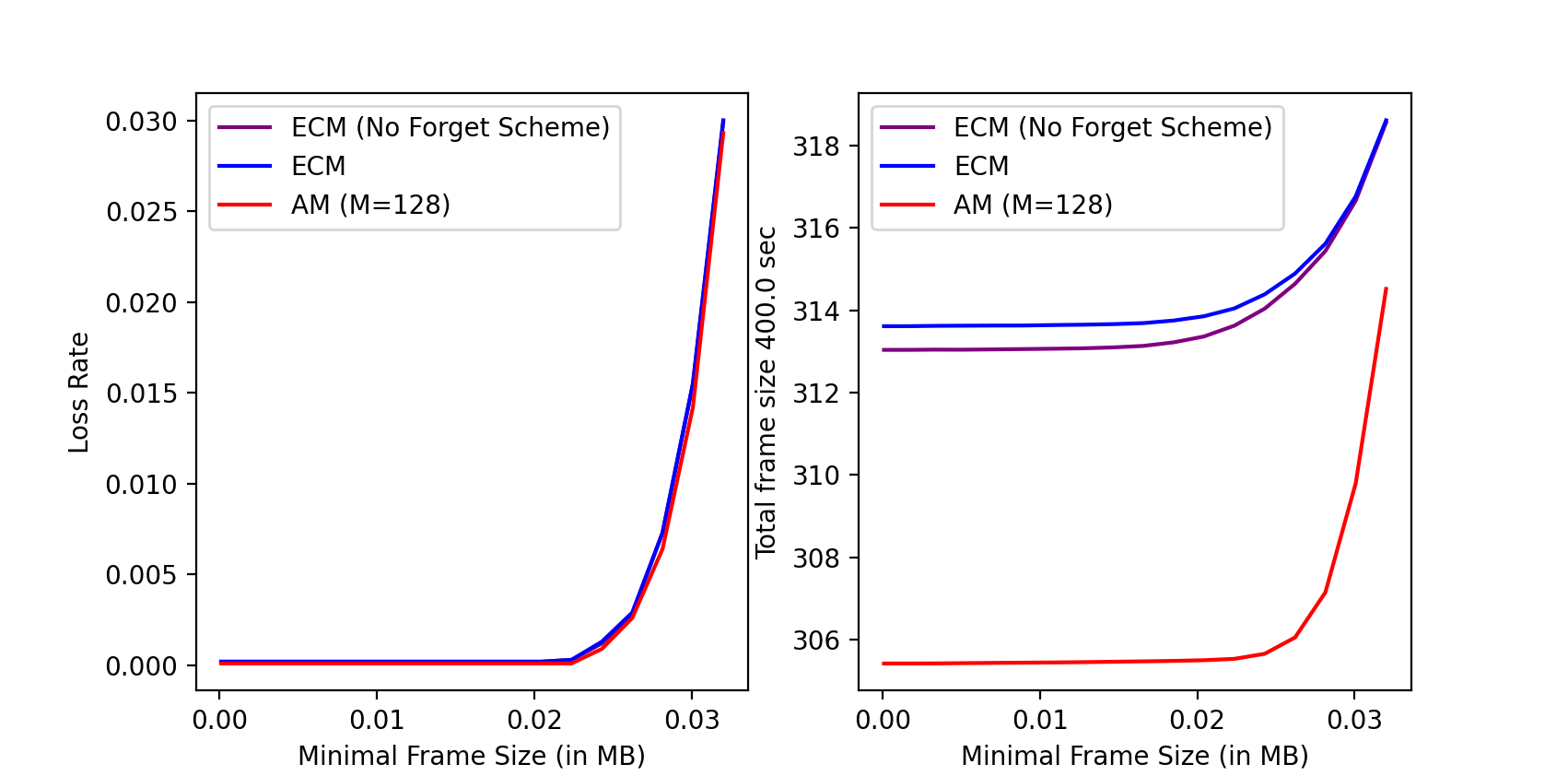}
	  \caption{Comparing with AM (M=128), ECM has similar loss rate and higher sent frame sizes.}
\end{figure}

The above figures (Figure 3, 4, and 5) show a typical result. 
We used a set of trace data under FPS equals 25 and let the testing period be at least 400 seconds to eliminate errors caused by selecting different testing sets. These three figures show a clear tendency that when the loss rate is
less than 2\%, 
ECM can send 2\% to 4\% more frame sizes out while having a similar or even less loss rate than AM. 
However, when FPS becomes more significant than 30, AM algorithms with large M perform better than ECM in the sum of frame sizes metric.

Based on the above experimental results, ECM outperforms AM ($M=1$) in most cases, 
where outperform means a higher sum of frames' sizes and a lower or similar loss rate. 
However, for $M=16$ and $M=128$ cases, 
ECM only performs better when the FPS is at a low level. 

This makes sense because when the FPS is high, i.e., 
the frequency of measuring the throughput becomes high, 
we collect many highly-correlated historical values frequently and put them into use directly in AM. 
These values' simply arithmetic average provides enough information about the future. 
While ECM requires a relatively massive data set of history measures to span its probability space, newly added data cannot immediately take effect.

On the other hand, when the FPS itself is low, the correlations between each measured $C_i$'s are intrinsically not that strong (because they are measured with a longer time in between); hence their mean becomes less informative, especially when $M$ increases. Thus ECM takes a turn to perform better in the long run. In this sense, ECM can be regarded as complementary for those higher-frequency measurement-based predicting algorithms.

\subsection*{Modeling the Condition Probability Distribution}

If we dig deeper into our Markov matrix $\mathcal{S_M}$, 
we may see what determines the accuracy of our ECM algorithm. 

\begin{figure}
	\includegraphics[width=0.48\textwidth]{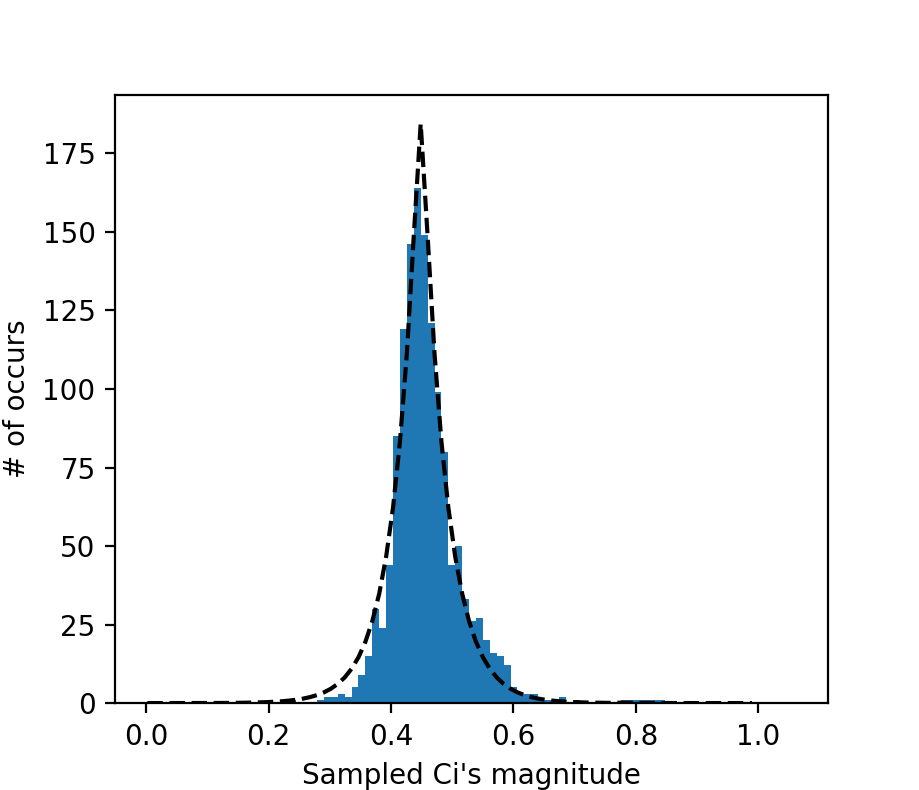}
	  \caption{An example of conditional probability/contingency distribution}
\end{figure}


Figure 6 shows a row slice taken from matrix $\mathcal{S}$, 
the contingency table. 
It shows what do those conditional probability/contingency distributions look like.
Checking by a goodness-of-fit test, 
Laplacian distribution happens to fit in this case. 
However, as the network environment changes, 
we cannot expect every conditional distribution to follow some regular pattern 
like the one shown in Figure 6. 
Figure 7 thus plots a row in another $\mathcal{S}$, 
which is way harder to model by some well-known mathematical configuration.

\begin{figure}[H]
	\includegraphics[width=0.48\textwidth]{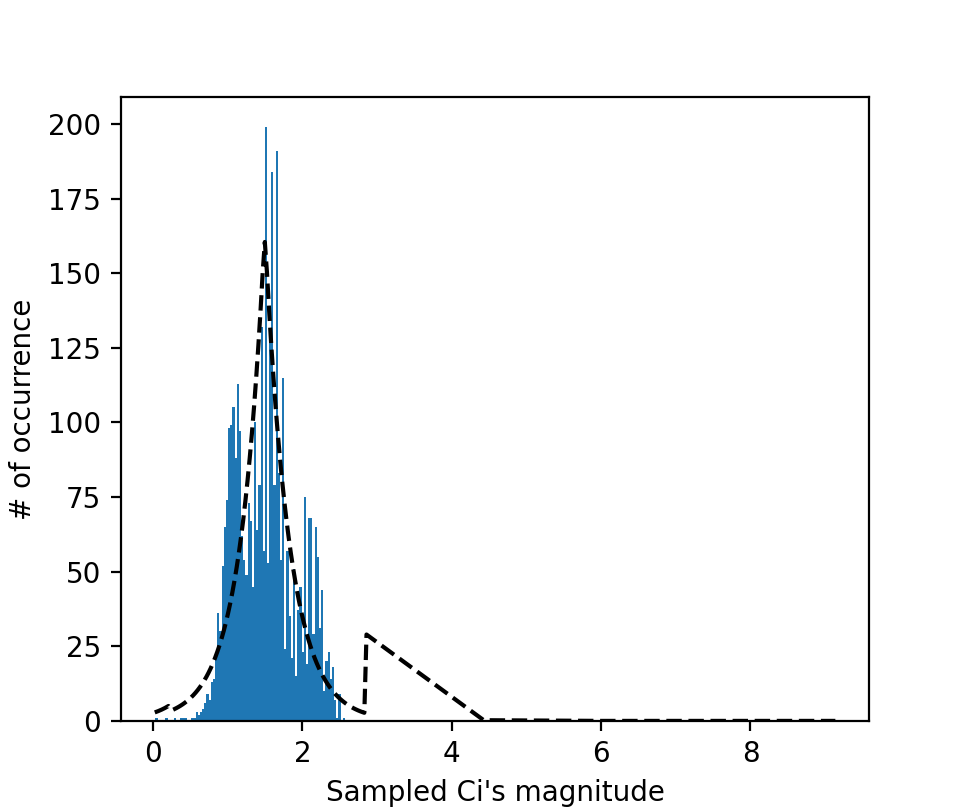}
	  \caption{Another example of conditional probability/contingency distribution}
\end{figure}

Recall that when FPS is lower i.e., 
each measurement of throughput value is farther from the other. 
Then conditional distributions of the $C_i$'s are more likely to 
behave like a recognized distribution with single peak and excess kurtosis.  
(E.g., Laplacian distribution shown in Figure 6) Empirically speaking, 
under those cases ECM can have a better performance.

\section{Conclusion}
This work proposed a new history-based method of 
predicting throughput, called ECM for its basing on empirical conditional probability 
by applying a Markov process assumptions to the changes of throughputs in mobile
networks. 

For estimating, we use the empirical conditional mean as the point estimator, 
and the empirical conditional probability for a confidence interval.

Moreover, we found that when the frequency of measuring network throughput (or FPS) is low, 
ECM outperforms AM by reducing the NRMSE for about $10\%$, 
and has a higher utilization rate of available bandwidth in our
video uploading simulator.

This implies that ECM may help live video uploaders with a limited network to equilibrium frame loss rate and video quality.

\end{document}